\documentclass[aps,prx,twocolumn,showpacs,psfig,superscriptaddress,longbibliography]{revtex4-1}

\usepackage{times}
\usepackage{graphicx}
\usepackage{float}
\usepackage{latexsym,bm,euscript}
\usepackage{color}
\usepackage{subfigure}
\usepackage{epstopdf}
\usepackage[colorlinks=true,linkcolor=blue,citecolor=blue]{hyperref}
\usepackage{hyperref}
\usepackage[normalem]{ulem}
\usepackage{mathrsfs}
\usepackage{amsmath,amssymb}
\usepackage{xspace}

\DeclareMathAlphabet{\mathpzc}{OT1}{pzc}{m}{it}

\newcommand{\citex}[1]{%
\begin{NoHyper}\citeauthor{#1} (\citeyear{#1})\,\end{NoHyper}\cite{#1}}

\newcommand{\Sec}[1]{Sec.~\ref{#1}}
\newcommand{\App}[1]{App.~\ref{#1}}
\newcommand{\Eq}[1]{Eq.~\eqref{#1}}

\newcommand{\EQ}[1]{Equation~\eqref{#1}}

\newcommand{\rsym}[1]{\ensuremath{\mathfrak{r}_{\mathrm{\ifx\@empty{}\else{#1}\fi}}}}

\def\QSpace{\mbox{QSpace}\xspace}
\def\onej{\ensuremath{1j}\xspace}

\begin{document}

\title{X-Symbols for Non-Abelian Symmetries in Tensor Networks}

\author{Andreas Weichselbaum}
\email{weichselbaum@bnl.gov}
\affiliation{Department of Condensed Matter Physics and Materials
Science, Brookhaven National Laboratory, Upton, NY 11973-5000, USA}
\affiliation{Physics Department,
Arnold Sommerfeld Center for Theoretical Physics, and Center for NanoScience,
Ludwig-Maximilians-Universit\"at, Theresienstrasse 37, 80333 Munich, Germany}

\begin{abstract}

The full exploitation of non-abelian symmetries in tensor
network states (TNS) derived from a given lattice Hamiltonian is
attractive in various aspects. From a theoretical perspective,
it can offer deep insights into the entanglement structure and
quantum information content of strongly correlated quantum
many-body states. From a practical perspective, it allows one
to push numerical efficiency by orders of magnitude. Physical
expectation values based on TNS require the full contraction of
a given tensor network, with the elementary ingredient being a
pairwise contraction. While well-established for no or just
abelian symmetries, this can become quickly extremely involved
and cumbersome for general non-abelian symmetries.
As shown in this work, however, the elementary step of a
pairwise contraction of tensors of arbitrary rank can be tackled
in a transparent and efficient manner by introducing so-called
X-symbols. These deal with the pairwise contraction of the
generalized underlying Clebsch-Gordan tensors (CGTs). They can
be computed deterministically once and for all, and hence they
can also be tabulated. Akin to $6j$-symbols, X-symbols are
generally much smaller than their constituting CGTs. In
applications, they solely affect the tensors of reduced matrix
elements, and therefore, once tabulated, allow one to completely
sidestep the explicit usage of CGTs, and thus to greatly
increase numerical efficiency.

\end{abstract}

\date{\today}

\maketitle

Tensor network states (TNS) provide a powerful natural framework
for the numerical treatment of strongly correlated quantum
many-body physics on lattice Hamiltonians \cite{Jordan08,
Jiang08, Orus09, Schollwoeck11, Orus19}. Starting in one
dimension (1D) with matrix product states (MPS), the numerical
renormalization group (NRG,
\cite{Wilson75,Wb07,Bulla08,Wb12_FDM})
and the density matrix renormalization group (DMRG,
\cite{White92,White04,Schollwoeck05,Stoudenmire12}) represent
powerful, non-perturbative, and accurate methods to deal with
strongly correlated system at arbitrary temperature both,
statically and dynamically. An attractive extension of the 1D
MPS structure was provided by the multiscale-entanglement
renormalization ansatz (MERA, \cite{Vidal07_MERA,Evenbly15}),
already also with an eye on higher dimensions, even if
significantly more expensive numerically. The very flexible
framework of TNS for lattice Hamiltonians has already also seen
a wide range of applications in two dimensions (2D) via
projected entangled pair states (PEPS)
\cite{Verstraete04_peps,Kraus10,Cirac09,Corboz10,Phien15} or
higher \cite{Xie12,Saez13,Ran17}, including applications in
quantum chemistry \cite{Nakatani13,Krumnow16,Orus19}. By
providing a natural algebraic structure to study entanglement
in strongly correlated systems, this also generated significant
interest from a quantum information perspective
\cite{Anders06,Schuch10,Eisert10} including symmetry protected
topological quantum phases
\cite{Levin05,Gu09b,Qi11,Hermele11,Capponi16}.
While highly efficient in 1D, the numerical cost for dealing
with TNS in 2D or higher, however, grows exorbitantly, even
though at least still polynomially. Therefore the exploitation
of symmetries is extremely relevant and important also on
practical grounds \cite{McCulloch02,Toth08,
Singh10,Alvarez10,Singh12,Moca12,Schmoll18,Nataf18}. In
particular, this applies for correlated systems in quasi-1D,
i.e., for long systems of narrow width, or for tree tensor
network states \cite{Murg10,Nakatani13,Singh13} in the presence
of multiple (symmetric) flavors in condensed matter systems
\cite{Costi09,Wb12_SUN,Hanl13,Stadler15,Nataf16,Toth10} or
optical lattices \cite{Greiner02,Bloch08,Cazalilla14,Scazza14}.

Symmetries on {\it all} indices in a tensor network state,
physical and bond indices alike, are well-defined only in 1D, or
more generally, in tree-tensor network (TTN) states
\cite{Murg10,Nakatani13,Singh13}, i.e, tensor network states
without loops along virtual bonds that link tensors. In the
presence of loops, significant ambiguities arise. Nevertheless,
on practical grounds, one typically sees that enforcing
symmetries on all indices also in a general TNS with loops shows
clear gains in efficiency \cite{Bauer11,Liu15}. An intuitive
handwaving argument for this may be provided based on the
interpretation of bonds as actual auxiliary state spaces which
motivated PEPS to start with \cite{Verstraete04_peps,Cirac09}.

In the presence of non-abelian symmetries
\cite{Dynkin47,Cahn84,LiE92,Pope06,Gilmore06,Humphreys12},
all tensors can be
decomposed into a tensor product structure of reduced matrix
element tensors (RMTs) and generalized Clebsch Gordan
coefficient tensors (CGTs) \cite{Singh10, Wb12_SUN} as a direct
generalization of the Wigner-Eckart theorem. This results in
two immediate consequences that can be used to greatly improve
numerical efficiency: (i) By splitting of a CGT factor for each
elementary, i.e., simple symmetry present, this allows one to
(strongly) reduce the effective dimensionality $D \to D^\ast$ of
a given index or state space by switching from a state-based
description to a description based on multiplets eventually
dealt with by the RMTs. (ii) The CGTs are purely related to the
symmetry of a given problem. Hence much of it can be dealt with
once and for all by tabulating the relevant information.

A general bottom-up framework for dealing with general
non-abelian symmetries in TNS was introduced in \citex{Wb12_SUN}
based on a general transparent tensor representation referred to
as \QSpace (cf. \App{app:QSpace:v3}). The approach taken
there was based on the explicit utilization, i.e., generation
and subsequent decomposition of CGTs. This is in stark contrast
to other approaches based on fusion trees, F-moves, etc., which
are essentially built on $6j$-symbols
\cite{Singh11phd,Schmoll18}. That approach works well for the
symmetry SU(2), where $6j$-symbols are readily available
analytically. But it becomes much more difficult for general
non-abelian symmetries such as SU($N>2$), SO($N>4$), or
Sp($2N>2)$ where due to the presence of outer multiplicity
$6j$-symbols are not at all readily available.

In contrast, the \QSpace approach is bottom up, as it solely
relies on the bare structure of the Lie algebra \cite{Wb12_SUN}.
The irreducible representations (ireps), multiplet fusion rules
and the corresponding CGTs thus generated were already all
tabulated in \cite{Wb12_SUN}.
However, contractions of CGTs were not tabulated due to the
presence of outer multiplicity (OM). The prescription in
\cite{Wb12_SUN} to deal with OM in the pairwise contraction of
tensors was to always recontract all CGTs based on their
particular instantiation in OM space. However, from a practical
point of view, this led to significant computational overhead
for larger non-abelian symmetries, such as SU($N\gtrsim4$).
As will be shown in the present work, however, there also exists
a transparent general way to deal with the problem of OM in the
pairwise contraction of CGTs based the introduction of so-called
X-symbols (where `X' is simply a reference to generalized tensor
multiplication, i.e., contraction). These can be computed once
and for all, and thus also be tabulated. X-symbols provide an
alternative approach to $6j$ symbols. Yet they are much more
naturally suited to tensor network algorithms, since they
strictly deal with the elementary operation of the contraction
of a pair of tensors (and hence a pair of CGTs) on an arbitrary
subset of shared indices.
 
Given the many reviews and detailed publications that already
exist on TNS, e.g., see \cite{Schollwoeck11, Wb12_SUN, Wb12_FDM,
Orus14, Orus19} and references therein, an elementary understanding
of tensor network states is assumed in this work. With this in
mind, the paper is organized as follows. \Sec{sec:TNS} sets the
stage with focus on symmetries in TNS which strongly builds on
\citex{Wb12_SUN}. \Sec{sec:prelim} provides conventions and
preliminaries required for the rest of the paper. \Sec{sec:ctr}
then introduces X-symbols and discusses their relevance in TNS,
followed by summary and outlook.

\section{Symmetries in tensor network states  \label{sec:TNS}}

Tensor network states typically describe lattice Hamiltonians,
with whom they share the same lattice structure. Whereas the
Hamiltonian may be longer ranged TNSs have nearest-neighbor
bonds only in order to minimize index loops. Each lattice site
$n$ is assigned a tensor $A_n$ that links its physical state
space $|\varphi_\sigma\rangle_{n}$ to its connected
(variationally determined) auxiliary state spaces $|a_x\rangle$.
In a pictorial language, the indices of a tensor are drawn as
lines, also referred to as the legs of a tensor. All state
spaces are indexed. E.g., the index $\sigma$ above spans the
local state space of a single physical site, whereas the index
$x=l,r,\ldots$ spans specific named bonds, such as l(eft),
r(ight), etc. [e.g. see \eqref{eq:indices}].

\subsection{Arrows on all legs}

The physical state spaces are orthonormal
from the very outset, having
${}_{n'}\langle \varphi_{\sigma'} | \varphi_{\sigma}\rangle_n=
\delta_{\ \sigma}^{\sigma'} \delta_{\ n}^{n'}$.
For generalized tree-tensor networks \cite{Murg10} including
matrix product states which, by definition, contain no loops
along any path of bond indices, cutting any auxiliary bond
separates the TNS into two disconnected blocks. As a direct
consequence, all auxiliary or bond state spaces can be made
orthonormal. If a given TTN is an exact symmetry eigenstate
globally, then all bond indices can be fully symmetrized, i.e.,
assigned symmetry labels without increasing the bond dimension
\cite{Singh13}. In the case of a TTN, these auxiliary bond
state spaces can be translated into well-defined orthonormal
effective quantum-many-body state spaces that represent entire
blocks of the system. Each such block only contains one open
bond index, starting from which it necessarily stretches all the
way to the open or infinite outer boundary of the physical
system, and hence also of the TTN considered.

For a TTN, many-body state spaces are typically generated
iteratively by adding one site after another to a block. This
way, by construction, any index or leg describes an orthonormal
many-body state space that either enters a tensor as part of a
tensor product space, or leaves a tensor with the interpretation
of an effective combined state space. So while several lines
may enter a tensor, there is always at most one line that can
leave a tensor. For finite TTN simulations then, auxiliary
indices all flow towards the orthogonality center (OC)
\cite{Stoudenmire10}, this is the only tensor in an entire TTN
that may have no outgoing index. It combines the orthonormal
states spaces of various blocks into a normalized global wave
function. If the global state is a singlet, this singleton
dimension may be skipped in which case the OC has no outgoing
index. If multiple global states are targeted, e.g., if the
global symmetry multiplet is not a scalar, then the OC also
needs to carry along an outgoing leg, namely the index that
resolves the global state or multiplet.

In a pictorial description, this naturally suggests that each
index (leg or line) in a tensor network (TN) is given an arrow.
Correspondingly, in mathematical notation, an outgoing
(incoming) index to a tensor can be written as a lowered
(raised) index, which is equivalent to covariant (contravariant)
index notation, respectively. A contracted, i.e., summed over
index then necessarily is outgoing from one tensor and ingoing
into another, consistent with Einstein summation convention that
an index is summed over if it appears twice, namely as a raised
and lowered index.

By having adopted the convention that the physical state space
of site $n$ is denoted by $|\varphi_\sigma\rangle_n$, i.e., with
lowered indices, this implies that the index (leg) in a tensor
where it enters must be a raised index. Thus combining, for
example, the state space of site $n$ with an effective bond
state space $a_l$ (`left'), the fused state space $a_r$
(`right') is given by,
\begin{equation}
\includegraphics[width=0.8\linewidth]{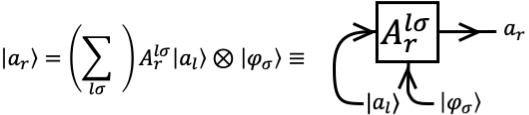}
\label{eq:indices}
\end{equation}
where the summation over double indices is implicit, hence the
bracket around the sum. TTNs guarantee that orthonormal state
spaces, and hence also symmetries, are well-defined throughout.
This is crucially important for efficient algorithms
\cite{Schollwoeck05,Schollwoeck11}, since these orthogonal state
spaces ensure an orthogonal environment, and hence optimal
conditions \cite{Lubasch14} when truncating bond dimensions to
the most entangled and thus most relevant quantum-many-body
states constituting a given global state. In this sense,
orthogonality and hence arrows are crucially important in any
TTN.

For a general TNS in the presence of loops, however, it is no
longer possible to associate all indices with well-defined
strictly orthonormal state spaces in the unique strong sense
this can be achieved with TTN \cite{Singh13}. Instead, one can
adopt the much less rigorous PEPS-like interpretation that bonds
host actual auxiliary state spaces where one simply imposes by
definition that these are orthonormal and adhering to the
symmetries of the overall Hamiltonian. When dynamically
truncating, and thus adapting auxiliary state spaces, it is no
longer possible then to have a perfectly orthonormal
environment. Instead, one needs to resort to optimal
conditioning \cite{Lubasch14}.
If one decides to exploit global symmetries, in any case, this
necessitates that symmetries must be enforced at every step in a
calculation. Here in particular, symmetries need to be enforced
locally with each tensor. As a consequence, all lines in a TNS
need to carry arrows. For TNSs with loops, such as a 2D PEPS,
certain individual tensors necessarily also need carry multiple
outgoing legs then. Independent of whether or not global
symmetries are enforced, however, when describing specific
algorithms in tensor networks, arrows on all legs, or
equivalently raised and lowered indices generally represent a
powerful natural concept for the underlying tensor algebra.

\subsection{General tensor decomposition}

In the presence of (non-abelian) symmetries, the tensor
coefficients $A$ (without hat) of any tensor operator $\hat{A}$
(with hat) can be decomposed into a sum over tensor products of
reduced matrix elements $\Vert A\Vert$ times generalized
Clebsch-Gordan coefficients $C$, \cite{Singh10,Wb12_SUN}
\begin{eqnarray}
   A = \bigoplus_q \Vert A\Vert_q \otimes C_q
\text{ .}\label{eq:tensor:decomp:0}
\end{eqnarray}
The tensor $A$ can have arbitrary rank $r$ which is defined here
as the number of legs (indices) attached. The sum over $q$
indicates sum over symmetry sectors. For each independent
elementary (i.e., simple) symmetry considered, another CGT can
be split off \cite{Wb12_SUN}, collectively written as $C_q$
above. To be precise, in this paper, a CGT always refers to a
specific elementary \mbox{(semi-)} simple symmetry such as
SU(N), Sp(2n), or SO(n) [see also \App{app:rsym}], and can
always also be chosen real. Yet the same way, an algorithm
needs to deal with RMTs and CGTs separately, also CGTs for
different symmetries can be dealt with completely separately and
in parallel. Hence while, for the sake of the argument for
simplicity only a single elementary non-abelian symmetry is
considered, all arguments in this paper straightforwardly also
generalize to the presence of multiple symmetries.

The decomposition in \Eq{eq:tensor:decomp:0} holds for general
symmetries, non-abelian and abelian symmetries alike. For
abelian symmetries, however, all multiplets contain only a
single state. Hence the tensorial structure of the CGTs reduces
to simple numbers there. For non-abelian symmetries, on the
other hand, one still also needs to systematically account for
outer multiplicity as explained in detail below. The overall
tensorial structure such as rank or directions of incoming and
outgoing indices, i.e., the arrow configuration, is exactly
inherited by all terms, RMTs as well as CGTs. 

In order to avoid excessive proliferation of indices, for the
sake of readability, simplified shortcut notations are adopted
as explained in the following. In particular, this concerns the
at times somewhat loose distinction between raised or lowered
indices if the distinction is not explicitly important in a
specific context. In \Eq{eq:tensor:decomp:0}, for example, the
combined symmetry labels for all legs have simply been written
as subscript $q$, even though legs may have mixed raised and
lowered indices.
The symmetry labels (simply also referred to as `$q$-labels')
for all $r=l+l'$ legs of a given CGT of rank-$r$ are given
by the set
\begin{eqnarray}
q &\equiv& (q^1\!, ..., q^{l}, q_{1'},..., q_{l'})
   \equiv  (q_1\!,... q_{l} \, | \, q_{1'},..., q_{l'})
\text{.} \quad \label{eq:qlabels}
\end{eqnarray}
Here each $q^i$ (or $q_{i'}$) is a tuple of multiplet labels of
fixed length that specifies the symmetry sector on a given leg
$i$ for all symmetries considered. While the order of legs
within the group of incoming (or outgoing) legs is important,
incoming and outgoing indices can be arbitrarily interspersed.
This means that the position of raised indices relative to
lowered indices is irrelevant. Therefore, e.g., all incoming
indices can be listed first, as shown in \Eq{eq:qlabels}. While
the first decomposition of $q$ in \Eq{eq:qlabels} explicitly
specifies raised and lowered indices, the equivalent last
decomposition splits the group of incoming ($l$ legs) via the
bar `$|$' from the group of outgoing indices ($l'$ legs). It
uses all lower indices, if relevant, otherwise. This is useful,
for example, when discussing standard rank-3 CGTs
$q\equiv (q_1 q_2|q_3)$ 
which fuse $q_1$ and $q_2$ into the combined total
$q_3$, also referred to as CG3 in this paper.
E.g., consider $(q \bar{q}|0)$ which fuses multiplet(s)
$q$ with their dual(s) into the scalar representation, simply
denoted as $0$. This has no more raised or lowered indices
whatsoever, which thus requires the last notation in
\Eq{eq:qlabels}.

Depending on the context, the combined label $q$ as a whole as
in \Eq{eq:tensor:decomp:0} may also be written as superscript,
instead, with no specific meaning of the location, unless
explicitly stated otherwise. Only when $q$ appears paired up as
raised and lowered index, e.g., $C_q D^q$, the legs $q$ are
considered traced over with arrows reverted in $D$ relative to
$C$ (cf. conjugate tensors below). Decomposed $q$ labels with
the same CGT always adhere to the interpretation of raised and
lowered indices.
The CGTs corresponding to the CG3s above then,
for example, correspond to $C_q \equiv C^{q_1 q_2}_{q_3}$ or
$C^{q \bar{q}}_0$. The same index convention as in
\Eq{eq:qlabels} also holds for the RMTs, i.e., using the
notation $n\equiv (n^1 \ldots n^{l} n_{1'},\ldots, n_{l'})$
for the index into the tensor of reduced matrix elements.
Here $n_i$ indexes individual symmetry multiplets in a given
symmetry sector $q_i$ in the state space decomposition of leg
$i$ as in $ |qn; q_z\rangle_i$ with $q_z$ the internal states
of a single multiplet in $q$ \cite{Wb12_SUN}. By convention,
in the notation of $q$ in \Eq{eq:qlabels} or also $n$, incoming
indices are listed first. Note, however, that this is reverse
to the bra-ket notation, e.g. as in \Eq{eq:indices}, which
places incoming states as kets, and hence needs to be read
right-to-left.

\subsection{Outer multiplicity\label{sec:OM}}

In the presence of non-abelian symmetries, tensors as in
\Eq{eq:tensor:decomp:0} are typically faced with the problem of
outer multiplicity (OM) in their CGTs. This means that for
exactly the same symmetry labels $q$ multiple ($m_q>1$)
orthogonal CGTs can arise [cf. \Eq{eq:Cnorm} later]. Hence the
CGTs acquire an additional multiplicity index $\mu = 1,\ldots,
m_q$, i.e., $C_q \to C_{q,\mu}$. If a given CGT has no outer
multiplicity, then $m_q=1$ which represents a singleton
dimensions and thus may be safely skipped. If a given
combination of symmetry labels is not permitted from a symmetry
point of view, then $m_q=0$. CGTs that are always free out
of outer multiplicity are CGTs of rank $r=0$ (scalars) or $r=2$,
as these are just proportional to the identity matrix.
Moreover, for rank $3$, CGTs that contain the defining irep or
its dual on one of their legs, i.e, {\it primitive} CGTs, are
also always OM free. Furthermore, CG3s 
$(q_1 q_2|q_3)$ where $q_3=q_1+q_2$ carries the maximum weight states
of $q_1$ and $q_2$ combined \cite{Wb12_SUN}
are also OM free since the maximum weight state is unique.

While there is no OM in the well-known SU(2) at the level of CG3s, 
OM routinely also occurs in SU(2) for CGTs of rank $r>3$ [see
\App{app:rsym}].  A simple example for the emergence of OM in SU(2)
for a rank-4 CGT is shown in (\ref{eq:OM}a),
\begin{equation}
\includegraphics[width=1\linewidth]{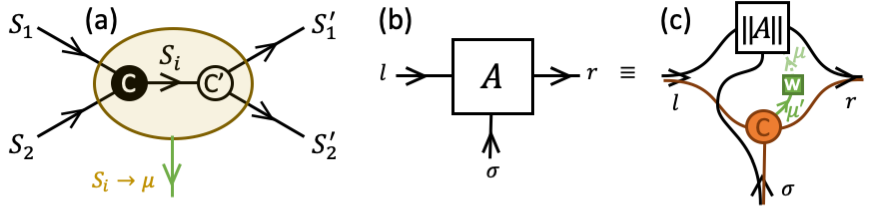}
\label{eq:OM}
\end{equation}
which represents the contraction of two multiplicity-free CGTs
$C$ and $C'$ of rank $3$ over the shared intermediate multiplet
$q_i=S_i$. However, typically there will exist multiple choices
for the intermediate spin multiplet $S_i$. Therefore when
`zooming out', i.e., contracting the intermediate index, the
resulting rank-4 tensor (brown circle) acquires outer
multiplicity $\mu=1, \ldots m_q$. Here $m_q$ is the number of
permitted intermediate $S_i$ for fixed open outer multiplets
$S_1^{(\prime)}$ and $S_2^{(\prime)}$. In the present case,
the presence of several permitted internal multiplets $S_i$
directly translates into OM indexed by $\mu$. Recoupling of the
internal (contracted) structure results in an orthogonal
rotation in OM space. For overall consistency, the OM index
[green line in (\ref{eq:OM}a)] also carries an arrow. Since
this comes `out' of a given CGT decomposition for $C \ast C'$,
arbitrarily but fixed, the index $\mu$ is chosen an outgoing
index in (\ref{eq:OM}a) [and of $C$ in (\ref{eq:OM}c)]. The
above point of view can be used to systematically determine the
level of OM for arbitrary symmetries and CGTs, as discussed
later in \Sec{sec:OM:dim}.

The presence of OM introduces an additional vector space for a
given CGT, referred to as outer multiplicity space. The CGTs in
\Eq{eq:tensor:decomp:0} then become a linear superposition in OM
space, described by the normalized coefficients $w_{\mu}^{\ \mu'} $,
\begin{eqnarray}
   A = \bigoplus_{q}
   \Bigl( 
   \Vert A\Vert_{q}^{\textcolor[rgb]{0.439,0.678,0.278}{\bf\mu}} \otimes \bigl(
      w_{ \textcolor[rgb]{0.439,0.678,0.278}{\bf\mu}}^{\ 
          \textcolor[rgb]{0.439,0.678,0.278}{\bf\mu'}}
      C_{q\textcolor[rgb]{0.439,0.678,0.278}{\bf\mu'}} \bigr)
   \Bigr)
\text{ ,}\label{eq:tensor:decomp}
\end{eqnarray}
with implicit regular summation over the multiplicity indices
$\mu$ and $\mu'$ (Einstein summation convention),
in contrast to the {\it direct} sum over symmetry sectors $q$.
Note, in particular, that the (block) summation over $q$ adds
to the overall dimension of the tensor $A$, whereas the
multiplicity indices no longer do (OM leads to additional multiplets,
which must have already been allocated at the level of the RMTs
$\Vert A\Vert_q$ \cite[e.g.\,see Fig.\,2 in][]{Wb12_SUN}).
All OM related indices are denoted in
green color consistent with the graphical notation as in
\Eq{eq:OM}. Effectively, the role of coefficient matrix $w$ is
such that it `ties' together \cite{Werner19}
the otherwise plain tensor product between RMTs $\Vert A\Vert_{q}$
and CGTs $C_q$, as graphically depicted in Eq.\,(\ref{eq:OM}c).

For a given symmetry sector $q$ in \Eq{eq:tensor:decomp}, $\Vert
A\Vert_{q}^{\mu}$ are the reduced matrix elements that come with
the CGT component $\tilde{C}_{q\mu}\equiv \sum_{\mu'} w_{\mu}^{\
\mu'} C_{q\mu'}$. In other words, for different orthogonal
outer multiplicity components, there can be a completely
different set of reduced matrix elements.
For this reason, the matrix $w$ in \Eq{eq:tensor:decomp} can
have at most as many rows (indexed by $\mu$) as there are
columns (indexed by $\mu' \leq m_q$). With all CGTs real, the
matrix $w$ can always be chosen also real and such that $w
w^T=1$. Fewer rows than $m_q$ are allowed in $w$ (the tensor
$A$ then simply does not span the full OM space), such that $w^T
w \neq 1$. However if there had been more than $m_q$ rows, QR
or singular value (SV) decomposition can be employed to
decompose $w = z \tilde{w}$. Contracting $z$ onto the RMT
$\Vert A\Vert_{q}^{\mu}$ in \Eq{eq:tensor:decomp}, $w$ can be
replaced now by the orthogonal $\tilde{w}$.

The CGTs $C_q$ can be computed once and for all and then stored
(tabulated) in a central, dynamically generated database. For
the storage of any tensor $A$ then only a {\it reference} to the
CGTs is required in terms of metadata. Associating $w$ with
$C_q$, i.e., considering $\sum_{\mu'} w_{\mu}^{\ \mu'}
C_{q\mu'}$, in principle, one can explicitly store these
individual OM components with $\Vert A\Vert_{q}^\mu$ for each
$\mu$, resulting in a non-unique listing of symmetry sectors $q$
in the listing of non-zero blocks of $A$ in
\Eq{eq:tensor:decomp}. This was the procedure in
\cite{Wb12_SUN}. However, instead, one can explicitly add an OM
index $\mu$ onto the RMT to the already existing indices $n$
\cite{Werner19}, and contract $w$ onto the RMT, instead, $\Vert
\tilde{A}\Vert_{q n}^{\mu'} \equiv \sum_\mu \Vert A\Vert_{q
n}^{\mu} w_{\mu}^{\ \mu'}$. Then the tensor $A$ can be written
as a {\it unique} listing over symmetry sectors $q$ [see
\App{app:QSpace:v3}]. While the
matrix $w$ may be eliminated this way, for practical purposes,
it may be explicitly kept with the tensor decomposition in
\Eq{eq:tensor:decomp} nevertheless. For example, this then
allows one to simply absorb operations in OM space such as
orthogonal transformations into $w$, while the likely much
larger CGTs and RMTs can remain unaltered.

\section{Conventions and preliminaries \label{sec:prelim}}

\subsection{Conjugate tensors}

Conjugate (or daggered) tensors arise whenever computing
expectation values $\langle \psi| .. |\psi\rangle \sim
\mathrm{tr}(A^\dagger .. A)$ or whenenver expressing terms in a
Hamiltonian that arise out of a scalar product, such as
spin-spin interactions, ${\bf \hat{S}}_i \cdot {\bf \hat{S}}_j
\sim S_i^\dagger \cdot S_j$ in between some sites $i$ and $j$.
In general, the {\it conjugate tensor} $A^\dagger$ of any given
tensor $A$ can be defined by the following prescription: take
the complex conjugate of all of its matrix elements, i.e.,
$A^\ast$ (relevant for RMTs only, since CGTs are real), then
revert all arrows, and take a mirror image of the tensor when
depicted graphically \cite{Wb12_SUN}. The latter is simply the
generalization of the transposition part in what one means by
$A^\dagger$, e.g. if $A$ is an operator sandwiched in between a
bra-ket bracket.
In terms of index order convention, taking a mirror image
implies w.r.t. the original tensor that if indices are read
clockwise in $A$ starting from the some arbitrary but fixed leg,
they need to be read counter-clockwise in $A^\dagger$. As
pointed out with \Eq{eq:qlabels}, though, incoming and outgoing
indices can be dealt with separately.

Taking the mirror image, actually, is a rather common graphical
procedure in TNS, even if one typically does not further dwell
on this. For example, when computing expectation values
$\langle\psi| ... |\psi \rangle$ for a given MPS, this involves
the contraction of an $A$-tensor as in \eqref{eq:indices} with
itself, as shown in (\ref{eq:Xconj}a),
\begin{equation}
   \includegraphics[width=1\linewidth]{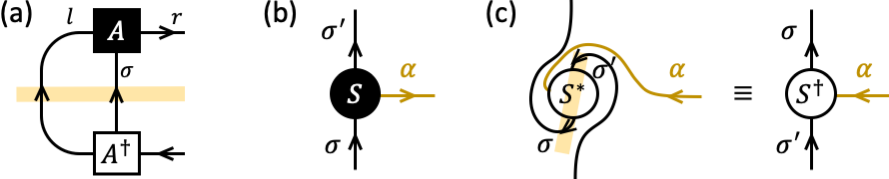}
\label{eq:Xconj}
\end{equation}
Here $A^\dagger$ (empty square at the bottom) is the mirror
image of $A$ (filled square at the top) with the arrows on all
lines reverted, while also taking complex conjugated matrix
elements. The mirror plane is indicated by the horizontal
yellow marker line. Similarly, the above procedure of
conjugating tensors also shows up systematically when dealing
with fermionic PEPS \cite{Corboz10} where in a pictorial
description $\langle \psi | .. |\psi\rangle$, the bra-state is a
full mirror image of the ket-state.

The other example already indicated is the representation of an
operator $\hat{S}$ that acts on a local state space of a site as
depicted in (\ref{eq:Xconj}b), e.g., which may be inserted into
the vertical line labeled by $\sigma$ in (\ref{eq:Xconj}a). In
general, the operator $\hat{S}\equiv \{ \hat{S}^\alpha \}$
consists of set of operators indexed by $\alpha$ which are
irreducible w.r.t. the symmetries under consideration. This
makes it an irreducible operator (irop), which thus naturally
acquires the third index $\alpha$ \cite{Wb12_SUN}. Non-scalar
irops are also non-Hermitian, e.g. considering fermionic
hopping, or spin raising and lowering operators.
Here (\ref{eq:Xconj}c) summarizes what one means by taking the
Hermitian conjugate of an operator: matrix elements are complex
conjugated ($S\to S^\ast$), top and bottom index are exchanged,
resulting in the tangled, yet non-crossing lines. Here also all
arrows are reverted, and a mirror image is taken w.r.t. to the
vertical yellow marker line which flips the irop index $\alpha$
to the left.
Untangling the lines by rotating the tensor clockwise by 180
degrees leaves the tensor unaltered, otherwise, and also does
not introduce any crossing of lines\cite{Corboz10}. This leads
to the final $S^\dagger$ in (\ref{eq:Xconj}c) which looks much
the same as (\ref{eq:Xconj}b), but now with reverted index
$\alpha$. This is required, e.g., when constructing scalar
products of operators as in $\hat{S}_i^\dagger \cdot \hat{S}_j$
where the dot product simply translates into a contracted
directed line indexed by $\alpha$.

This paper also adopts the graphical convention that tensor
conjugation switches from a filled tensor (box, circle, etc.) to
an empty outlined object, or vice versa, as already seen in
\Eq{eq:Xconj}. Similarly, in (\ref{eq:OM}a) the CGT $C$ 
with two incoming legs was depicted by a filled circle,
in contrast to the conjugate tensor $C'$ with two outgoing legs
(empty circle). In a mathematical notation, finally, tensor
conjugation can be denoted by raising or lowering its indices,
e.g., $\bigl( C_{q \mu} \bigr) ^\dagger \equiv C^{q \mu}$.

\subsection{CGT normalization convention in OM space \label{sec:CGT:norm}}

When considering a CGT $C_q$ for an arbitrary but fixed
set of symmetry labels $q$, a convenient normalization
convention in OM space adopted throughout is,
\begin{eqnarray}
   \mathrm{Tr}\bigl( C_{q \mu} C_{q \mu'}^\dagger \bigr) \equiv
   \mathrm{Tr}\bigl( C_{q \mu} C^{q \mu'} \bigr)  = 
   \delta_{\mu}^{\ \mu'} 
\text{ ,}\label{eq:Cnorm}
\end{eqnarray}
where $\mathrm{Tr}\bigl( C_{q \mu} C_{q \mu'}^\dagger \bigr)$
stands for the full contraction (tensor trace) of $C_q$ with the
conjugate of itself, only keeping OM indices open and having
fixed $q$. By pairing with its conjugate tensor, this also
reverts the arrow of the OM index, hence raises the index $\mu'$
in \Eq{eq:Cnorm}. The CGT normalization in \Eq{eq:Cnorm} simply
generates an orthonormal basis in the OM `vector space' which is
convenient when performing explicit decompositions or
projections in OM space.
\EQ{eq:Cnorm} fixes the CGT $C_q$ up to an orthogonal
transformation in OM space (cf. matrix $w$ in \Sec{sec:OM}). As
the CGT will be generated dynamically depending on the
calculation, this makes it history dependent, adding OM
components as they occur \cite{Wb12_SUN}. This fixes $C_q$ up to
an overall sign convention. Including $\mu$ as last index in
$C_q$ and assuming column-major index order, the standard sign
convention of Clebsch-Gordan coefficients is followed, namely
that the first non-zero matrix element of the full CGT $C_q$ is
chosen positive. Therefore the component for $\mu=1$ fixes the
sign of the entire $C_q$ for $\mu\leq m_q$.

For rank-3 CGTs, the normalization in \Eq{eq:Cnorm} is closely
related to the normalization of Wigner-3j symbols \cite{Messiah66}.
As such, it is different from the normalization of the standard
Clebsch-Gordan coefficients e.g. for SU(2) $(S_1 S_2 | S_3)$
that fuse $(S_1,S_2)$ into an orthonormal basis in $S_3$ (and
not in OM). The latter would result in a norm in \Eq{eq:Cnorm}
that is equal to $|S_3|$, i.e., the dimension of multiplet
$S_3$. Consequently, this would make the normalization
dependent on the direction of arrows in a given CGT. This is
rather inconvenient on general grounds, and specifically so in
the TNS context, since the presence or not of OM {\it does
neither depend on the direction of arrows nor on the specific
order} of symmetry labels.
This simply follows from the discussion below that raising or
lowering of an index on a CGT is equivalent to applying a
unitary matrix on that index while also switching to the dual
representation on that leg [cf. \Sec{sec:1j}]. Conversely, a
permutation of indices on a CGT can only induce orthogonal
rotations in OM space [cf. \Sec{sec:perm}] which clearly also
leaves the OM dimension invariant.

Therefore based on the normalization in \Eq{eq:Cnorm}, the CG3s 
$q=(q_1,q_2 | q_3) \equiv (q^1 q^2 q_3)$ have
the same normalization for any permutation of the symmetry
labels together with arbitrary raising or lowering of indices.
Raising and lowering of labels, however, changes the
interpretation of which multiplets are fused together. By
raising all indices in a CGT, for $q=(q^1 q^2 q^3)$, this is
equivalent to fusing all three multiplets of a CG3 
into a total scalar representation denoted by $q_\mathrm{tot}=0$,
i.e. $q=(q_1,q_2, q_3|0)$, and subsequently skipping the
trailing singleton dimension. The resulting rank-3 CGT has
all-incoming legs, i.e. $q=(q^1 q^2 q^3)$, which is exactly
equivalent to the Wigner-$3j$ (here `$3q$') symbol up to an
overall sign.

\subsection{Fully contracted CGT networks \label{sec:Tr:full}}

When a tensor network is fully contracted, all indices are
paired up and summed over. This holds for both, the RMTs as well
as the CGTs. Here, for the sake of the argument, however, the
focus is on an isolated fully contracted TNS solely comprised of
CGTs. The OM for each participating CGT is assumed fixed to
some arbitrary but fixed linear superpositions in OM space [e.g.
see \Eq{eq:tensor:decomp}], such that there is no open index
left in the TNS, and the full contraction yields some number
$x$. By construction, this number can be non-zero only if the
CGT network is permitted from a symmetry point of view. Now if
one opens up a single leg in this otherwise fully contraction
CGT network, say with symmetry label $q_i$, the result is
proportional to a rank-2 CGT $C_{q_i}^{q_i}$, i.e., with one
incoming and one outgoing index. The only such CGT that exists
is the identity matrix up to normalization. Its graphical
representation is a single directed line. Therefore it follows
that a CGT network that is fully contracted up to a single
opened up index $i$, necessarily is proportional to the identity
matrix $1^{|q_i|}$ of dimension $|q_i|$, i.e., the size of
multiplet $q_i$.

As a specific example, consider the tensor network in \Eq{eq:Cnorm}
of just one CGT fully contracted with itself for arbitrary but
fixed $\mu$ and $\mu'$. Then opening up one bond index $i$
with symmetry label $q_{i}$, one obtains
\begin{eqnarray}
   \mathrm{Tr}_{\backslash q_i}\bigl( C_{q \mu} C_{q \mu'}^\dagger \bigr) =
   \tfrac{\delta_{\mu}^{\ \mu'}}{|q_i|} 1^{|q_i|}
\text{ ,}\label{eq:Cnorm:1}
\end{eqnarray}
The normalization is determined by the requirement that the
final trace over $i$, when performed, results back in
\Eq{eq:Cnorm}.

\subsection{Reverting arrows and \onej-symbols \label{sec:1j}}

Reverting the arrow on a given bond in a TNS changes its
interpretation, as well as the interpretation of the associated
tensors. As this will be useful also later in the CGT context,
consider first the elementary process where the orthogonality
center in an MPS is iteratively propagated from site $n \to
n{+}1$ with associated tensors $\tilde{A}_n$ and $A_{n+1}$,
e.g., see \Eq{eq:indices}, contracted on their shared auxiliary
bond \cite{Schollwoeck11}, 
\begin{eqnarray}
\tilde{A}_n A_{n+1}
&=& \bigl( A_n \underbrace{\tilde{X}_n\bigr) A_{n+1}}_{
  \equiv \tilde{A}_{n+1}}
\equiv
  \underbrace{\tilde{A}_n \bigl( \tilde{X}_n^{-1}}_{
  \equiv A_n } \underbrace{\tilde{X}_n \bigr) A_{n+1}}_{
  \equiv \tilde{A}_{n+1}}
\text{ ,}\label{eq:bupdate}
\end{eqnarray}
where the tilde indicates the tensor that carries the OC.
Before the iteration step, the OC is located on site $n$, and
the bond in between sites $(n,n+1)$ describes an orthonormal
effective many-body state space for the entire right block of
sites $n'>n$. After the iteration step, conversely, the arrow
on the bond changed its direction, and now describes the
orthonormal effective many-body space for the entire left block
of sites $n' \leq n$. On a procedural level, one starts with
the tensor $\tilde{A}_n$ in \Eq{eq:bupdate} that carries the OC
and performs QR or SV decomposition on it
\cite{Schollwoeck11,Wb12_SUN}. This yields $\tilde{A}_n = A_n
\tilde{X}_n$, where $A_n$ is a new isometry, and the OC is now
shifted onto the tensor $\tilde{X}_n$ located on the bond in
between sites $(n,n+1)$. When contracted onto $A_{n+1}$, this
makes the former isometry $A_{n+1}$ the new OC
$\tilde{A}_{n+1}$.
Formally, in the last equality of \Eq{eq:bupdate}, the direction
of a bond is flipped by using the Gauge freedom inherent to TNS
\cite{Schollwoeck11}. This consists of inserting the identity
$1 = \tilde{X}_n^{-1} \tilde{X}_n$ and then associating the
tensors $\tilde{X}_n^{-1}$ and $\tilde{X}_n$ with the left and
right $A$-tensors, respectively. However, as seen from the left
of \Eq{eq:bupdate}, it is not necessary to actually compute the
inverse of $\tilde{X}_n$ here as this may be ill-conditioned.
Overall, one can exactly flip the orientation of a leg in a TNS
without changing the global physical state. One only changes the
local perspective and interpretation by splitting off the tensor
$\tilde{X}_n$ from $\tilde{A}_n$ and contracting it onto the
neighboring tensor $A_{n+1}$.

Now the discussion of flipping the direction of an arrow an a
given elementary CGT $C_q$ with fixed symmetry labels $q$
follows much of the same spirit. In contrast to RMTs, however,
the procedure is naturally much more constrained for CGTs. For
example, a CGT is already always in a canonical form [cf.
higher-order SV decomposition, \onlinecite{Xie12}]. Even more,
with \Eq{eq:Cnorm:1}, it has constant singular value spectrum
w.r.t. to any of its bonds, since by symmetry, all states in a
given multiplet are necessarily equally important. As will be
shown below then, the matrix $\tilde{X}$ in \Eq{eq:bupdate} to
flip the direction of an arrow must be unitary for CGTs, while
in the same process the irep label $q$ also needs to be switched
to its dual $\bar{q}$.

In order to show this, it is sufficient to narrow the discussion
further down to a single directed line with symmetry label $q$,
assuming a single elementary non-abelian symmetry for the sake
of the argument without restricting the case. This line may
represent, e.g., an auxiliary bond in a TNS, and may be
associated with the CGT $C_q^{\ q} \propto 1^{|q|}$.
In general, now irep $q$ has a unique dual representation
$\bar{q}$. This dual shares the same multiplet dimension, i.e,
$|q| = |\bar{q}|$, and it is the only irep that, when fused
together with $q$, permits the scalar representation as an
outcome, i.e., having $(q \bar{q} | 0)$ exactly once. The
corresponding CGT $C_0^{q\bar{q}}$ therefore never has OM, but
is unique up to an overall sign convention.
By definition, the scalar representation, always denoted by the
label $0$ here, is fully symmetric under symmetry operations.
Hence its multiplet only consists of a single state, i.e.,
$|0|=1$. For example, SU(2) is self-dual, i.e., $\bar{q}=q$ for
all $q$, such that the product space of spin $q=S$ with itself
(and itself only) always also yields a singlet with
$S_\mathrm{tot}=0$, having $(SS|0)$. The same argument can be
further extended also to U(1) abelian symmetries such as charge
($N$) or spin ($S_z$). There the dual is simply given by
$\bar{q}=-q$, since $q+(-q)=0$.

By making use of the dual representation, this allows one to
define the unitary matrix $U^{q \bar{q}}_{(0)} \equiv \sqrt{|q|}
\cdot C_0^{q\bar{q}}$ of dimension $|q|$ for any representation
$q$ as depicted in (\ref{eq:revert}a),
\begin{equation}
   \includegraphics[width=1\linewidth]{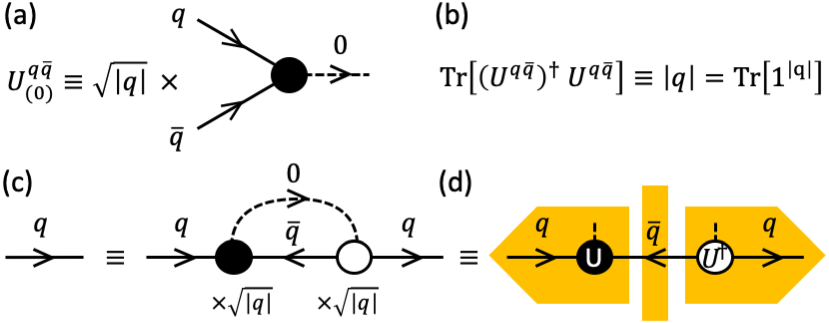}
\label{eq:revert}
\end{equation}
The singleton dimension in the scalar representation, indicated
by the dashed line to the right, will be frequently skipped in
$U$ to emphasize that $U$ is a matrix, indeed. Hence the
subscript $0$ has been written in brackets. Now
$\mathrm{Tr}(U^\dagger U)$ [Eq.\,(\ref{eq:revert}b)] is
proportional to a rank-3 CGT fully contracted with its
conjugate. So when opening up a single index, namely the one
for $q$, this must be proportional to the identity matrix [cf.
\Eq{eq:Cnorm:1}], and with the normalization as chosen in
Eq.\,(\ref{eq:revert}a), $U^{q\bar{q}}$ must therefore be
unitary (strictly speaking, orthogonal, since CGTs are always
real).

This now permits to insert a simple identity $U^\dagger U {=}1$
on a given leg in a TNS, as depicted in Eq.\,(\ref{eq:revert}c)
which provides a simple recipe to revert the arrow on any
contracted line with symmetry label $q$ that links CGTs, much in
the spirit of \Eq{eq:bupdate} earlier. By simply contracting
$U$ and $U^{T}$ onto the two neighboring tensors linked via the
contraction, as indicated by the broad arrows to the left and to
the right in light color in (\ref{eq:revert}d), the arrow on the
black line in the center now points in the opposite direction as
compared to (\ref{eq:revert}c). At the same time, the symmetry
label on the bond flipped to its dual, i.e., $q\to \bar{q}$.
When not mentioned explicitly, the latter will be implicitly
assumed whenever individual indices are lowered or raised. Since
the unitary matrix $U^{q \bar{q}}$ effectively only refers to a
single symmetry label $q$, the unitary $U^{q\bar{q}}$ is
referred to as a \onej-symbol with reference to the literature
on SU(2) \cite{Derome65,Butler75}.

One needs to be careful, however, when skipping the singleton
dimension [dashed line in (\ref{eq:revert}c)], which is reduced
to the little residual `stems' to the top in (\ref{eq:revert}d):
these stems are important to keep track of the order of
$U^{q\bar{q}}$ vs. $U^{\bar{q} q}$ e.g., for self-dual $q$.
They must point in the same direction, to ensure that the
conjugate of precisely the {\it same} object $U^{q\bar{q}}$ is
inserted together with $U^{q\bar{q}}$ in order to guarantee an
identity matrix overall. Specifically, the dashed line in
(\ref{eq:revert}c) must not cross the solid horizontal line.
Otherwise sign errors can arise, since $[U^{q\bar{q}}]^T = \pm
U^{\bar{q} q}$ where the sign depends on $q$. For example, for
SU(2) a sign arises for all half-integer spins.

In summary, \onej symbols can be utilized to revert arrows on
lines in a TNS, or equivalently, to raise or lower indices in
mathematical notation (in this sense, the \onej symbol acts like
a metric tensor within the tensor algebra of a given multiplet).
In principle, therefore it also suffices to tabulate the CGTs
with all-upper indices only (all incoming), since indices can be
simply lowered by applying \onej's.

\subsubsection{Generation of \onej-symbols}

A \onej-symbol can be computed, obviously, via an irep
decomposition of $(q \bar{q}|\ast)$. Starting from maximum
weight states, however, the largest ireps are always generated
first, with the \onej-symbol the very last CGT to be generated.
This is not practical for very large multiplets, bearing in mind
that $|q| = |\bar{q}|$. Also, in TNS one typically fuses a
given large effective state space (bond index) with new local
state spaces of small dimension. That is large (effective)
multiplets get fused routinely with (much) smaller ones. But for
most part, one can avoid fusing two large multiplets. This is
specifically important for large symmetries such as SU($N\gtrsim
4$) [cf. \App{app:rsym}].

Therefore an alternative route to computing \onej-symbols is
desirable. Note that \onej-symbols are only square matrices of
dimension $|q|$, which is in stark contrast to fully decompose a
$|q|^2$ dimensional vector space into irreducible
representations. Moreover, \onej-symbols are (close to)
anti-diagonal, i.e, very sparse like CGTs in general. In the
absence of inner multiplicity (IM) \cite{Wb12_SUN} such as for
SU(2), they are strictly antidiagonal with alternating entries
$\pm 1$.

The non-trivial part for \onej-symbols arises from the
anti-diagonal block structure in the presence of IM which
requires consistent conventions on how to decompose IM spaces
\cite{Wb12_SUN}. The \onej symbol derives from a CG3. 
When the underlying symmetry already permits OM for CG3s, 
this also implies the presence of IM, and
hence block structure in \onej symbols. The \onej symbol
itself, however, is unique otherwise up to a global sign
convention which is simply inherited here from the overall sign
convention on CGTs, as discussed with \Eq{eq:Cnorm}.

The approach taken in \QSpace [cf. \App{app:QSpace:v3}] then to
compute \onej symbols is based on the fact that the scalar
multiplet $q=0$ is destroyed by every one of the
$\alpha=1,\ldots,\rsym{sym}$ generalized raising and lowering
operators $S_\alpha^{(\dagger)}$\!, with $\rsym{sym}$ the rank
of a given symmetry [cf. \App{app:rsym}; note that $\rsym{sym}$
needs to be differentiated from the rank $r$ of a tensor
which is just the number of its legs].
Having explicit
access to the sparse generators $S_\alpha^{(\dagger)}$ in the
representation $q$ of some given symmetry of rank $\rsym{sym}$,
one can resort to a variational Krylov based minimization, and
compute the ground state of the sparse pseudo Hamiltonian (cost
function),
$$
   H_{q}^{1j} = \sum_{\alpha\leq \rsym{sym}}
   \bigl( S_\alpha^{\phantom{\dagger}} S_\alpha^{\dagger}
   + S_\alpha^{\dagger} S_\alpha^{\phantom{\dagger}}
   \bigr)_q \text{ .}
$$
This is in general a well-conditioned problem with a unique
ground state (the \onej-symbol) at `energy' zero and with a
`gap' of order 1. By construction, this ground state must be
simultaneously maximum {\it and} minimum weight state, hence
represents a scalar multiplet. The Kryolv based minimization
then allows one to directly converge the \onej-symbols via
iterative means down to ones numerical floating point precision.
The sparse nature of the the \onej-symbols strongly limits the
variational parameter space, and hence leads to fast
convergence.

\subsubsection{\onej-symbols via contractions}

A useful application of \onej-symbols arises when computing a
ground state of a system which itself is in a global singlet
symmetry sector, i.e., the scalar representation
$q_\mathrm{tot}=0$. By skipping this global singleton
dimensions, this requires access to simple \onej-symbols during
setup. However, when sweeping through the TNS, the OC repeatedly
gets located on an auxiliary bond deep inside the TNS where a
wide range of multiplets can be explored. Here the shifting of
the OC can also be achieved by contraction, e.g., by projecting
onto identity $A$-tensors (which correspond to simple $A$-tensor
as in \Eq{eq:indices} yet initially without truncation
\cite{Wb12_SUN}). The tensor $\tilde{X}$ that carries the OC
onto a bond then is of rank-2 with all indices incoming.
Therefore, up to normalization, all of its CGTs necessarily must
correspond to \onej-symbols. In this sense, \onej-symbols can
also be generated via contractions.

\subsection{Determination of OM dimension \label{sec:OM:dim}}

Outer multiplicity of a given CGT is independent of the
direction of its legs. This is apparent from the above explicit
construction of reverting arrows which solely corresponds to
applying a specific unitary on a given leg. Therefore OM is an
intrinsic quantity of a CGT. Consequently, the OM index in a
pictorial description needs to be attached to the tensor itself
(and not to any of its legs), as already shown with the tensor
$C$ in Eq.\,(\ref{eq:OM}c).

The OM dimension for rank-3 CGTs can be determined via standard
fusion rules of a pair of irreducible representations. If such
a tensor product arises from building a quantum many body state
starting from the vacuum state and iterative fusion of local
state spaces, they need to be computed in full via a standard
decomposition of a pair of ireps \cite{Wb12_SUN}. These CG3s 
also build the elementary basis and starting point for
subsequent contractions of CGTs.

Now if one encounters a CGT of rank $r>3$, its full OM can be
determined iteratively in a constructive way from smaller rank
tensors assuming that their OM is known. For example, for a
rank-4 CGT, the example in (\ref{eq:OM}a) for SU(2) can
generalized as follows: since arrow directions can be altered at
will without affecting the OM dimension (bearing in mind to also
switch to dual representations), one can build the sequential
MPS-like structure in (\ref{fg:OM:dim}a),
\begin{equation}
   \includegraphics[width=1\linewidth]{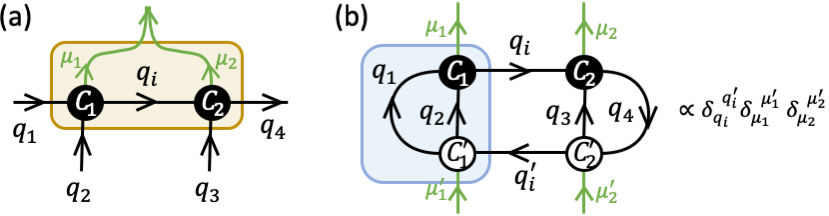}
\label{fg:OM:dim}
\end{equation}
by taking the fused multiplet $q_i$ out of a CGT $C_1$ into a
tensor product space with irep $q_3$, and then picking the
desired irep $q_4$. For fixed $q$'s then, given that there is
no loop in (\ref{fg:OM:dim}a), the combined OM $\mu_\mathrm{tot}
= 1, \ldots,m_\mathrm{tot}$ of the CGT described by brown box is
simply the product of the multiplicities of $C_1$ and $C_2$.
Bearing in mind, that the intermediate contracted multiplet
$q_i$ can vary, the total OM of the CGT with $q=(q_1 q_2 q_3
q^4)$ is given by,
\begin{eqnarray} 
   m_\mathrm{tot} = \sum_{q_i} m_1(q_i) \times m_2(q_i)
\text{ .}\label{eq:OM:dim}
\end{eqnarray} 
This can be shown by building an OM basis for the CGTs derived
from (\ref{fg:OM:dim}a) for all $\mu_1 {\leq} m_1$, $\mu_2
{\leq} m_2$, and for all permitted $q_i$. However, these are
already all orthogonal to each other, as seen from computing
their overlap as in (\ref{fg:OM:dim}b). Starting with the
orthogonality of CG3s, the contraction in
left blue box in (\ref{fg:OM:dim}b) is proportional to the
identity matrix, i.e., reduces to a simple line with weight
$\propto \delta_{\mu_1\mu'_1} \delta_{q_i q'_i}$. When repeated
iteratively with the next (here last) pair of CGTs $C_2$ and
$C'_2$, this directly leads to \Eq{eq:OM:dim}. As emphasized
there, the multiplicities $m_1$ and $m_2$ clearly depend on the
choice of $q_i$.
The above procedure can be extended towards any sequence of CGTs
also of higher rank, that are contracted in a linear sequence
without loops.
As a result this demonstrates, that the full outer multiplicity
space grows rapidly (exponentially) with increasing rank of a
tensor.

As a general strategy then to avoid proliferation of OM spaces,
this suggests (i) to reduce the rank of a tensor by fusing
indices as far as possible in a TN algorithm. Moreover, the
actual level of OM generated also depends on the specific TN
calculation performed. For the largest CGTs encountered,
typically a far smaller OM space is explicitly generated by
contractions than theoretically possible. Hence (ii) one can
refrain from insisting to build the full OM space in any
circumstances encountered. Rather, one can build the OM space on
demand [see \App{app:QSpace:v3}]. If a new OM component is
encountered via contractions, it can be added once and for all
to ones database. On the downside, a build up of the CGT
database this way becomes dependent on the history of
calculations. So one must be extremely careful to ensure
consistency across independent calculations or threads that
simultaneously access the same central database. This can be
achieved by coordinating updates, e.g., via locking mechanisms,
in order to avoid race conditions resulting in inconsistent
histories.

\section{Contractions \label{sec:ctr}}

\subsection{Pairwise contractions and X-symbols}

Contractions in any TN state are always tackled by elementary
pairwise contractions, in practice,in complete analogy to
evaluating the product of multiple matrices. Hence the
elementary step for contracting a TN state is the contraction of
two tensors. To be specific, consider some rank-4 tensor with
arbitrary but fixed index order $1,\ldots,4$, as shown in
(\ref{fg:tensor4}a),
\begin{equation}
   \includegraphics[width=1\linewidth]{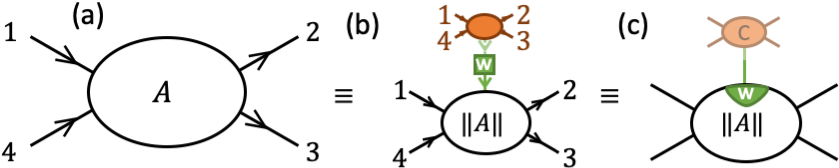}
\label{fg:tensor4}
\end{equation}
In the presence of symmetries, it is redrawn schematically in
the spirit of (\ref{eq:OM}a,c) in (\ref{fg:tensor4}b), depicting
the decomposition of symmetry sectors into the tensor product of
RMTs and CGTs. For simplicity only a single CGT is shown, while
in the presence of multiple symmetries, each has its own CGT.
Also with reference to \Eq{eq:tensor:decomp}, there is a sum
over symmetry sectors while in the pictorial representation in
(\ref{fg:tensor4}b) the focus is on one arbitrary but fixed set
of symmetry labels $q$ with the matrix $w$ that links the CGT to
the corresponding RMT. To further simplify the following
discussion, (\ref{fg:tensor4}b) is redrawn in
(\ref{fg:tensor4}c) with indices and arrows removed, while
bearing in mind that, of course, arrows and index order stay
intact. Also the matrix $(w_q)_\mu^{\ \mu'}$ can be fully
merged with (i.e., contracted onto) $\Vert A\Vert_q$. Hence the
matrix $w$ is not explicitly needed for the sake of the argument
here, and thus is skipped. The shading of the CGT at the top of
(\ref{fg:tensor4}c), finally, indicates that the CGT itself does
not need to be explicitly stored with the tensor $A$ itself, but
that a reference to a central database suffices. This way the
multiplicity index with the RMT automatically represents an open
index, as suggested by the grouping in \Eq{eq:tensor:decomp}.
Depending on the contraction, however,
the resulting CGTs may need to be updated centrally, e.g., if a
new OM component is encountered.

Now consider a pair of tensors, $A$ and $B$, as in
\eqref{fg:tensor4} contracted on a shared set of indices (state
spaces), as shown in (\ref{fg:xsym1}a),
\begin{equation}
   \includegraphics[width=1\linewidth]{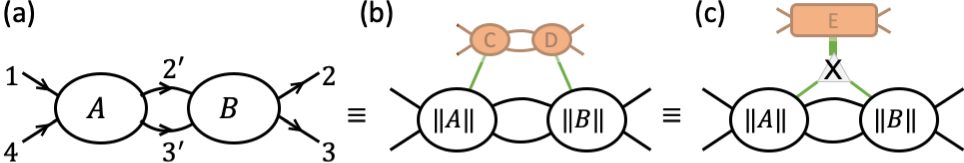}
\label{fg:xsym1}
\end{equation}
This is a generalized matrix multiplication, thus is
symbolically written as $A\ast B$. In the presence of
non-abelian symmetries, they contain references to CGTs
[(\ref{fg:xsym1}b)]. Via the tensor product structure in
\Eq{eq:tensor:decomp}, the tensorial structure is exactly the
same for both, RMTs as well as CGTs. Therefore when performing
a contraction of two tensors on a specified set of indices in a
TNS, precisely the same contraction needs to be performed on the
level of the RMTs, $\Vert A\Vert \ast \Vert B \Vert$, as well as
on the level of CGTs, $C\ast D$. These are separate from each
other, and hence can be dealt with completely independently.
The contraction of the RMTs always needs to be performed
explicitly, as this is part of the physical problem under
investigation. The contraction of CGTs, however, is purely
related to symmetries, and hence can be computed once and for
all and tabulated. Now consider the contraction of some
arbitrary but fixed pair of CGTs, $C \ast D$ as in
(\ref{fg:xsym1}b), shown in (\ref{fg:xsym2}a),
\begin{equation}
   \includegraphics[width=0.96\linewidth]{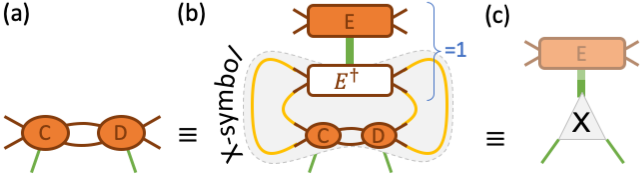}
\label{fg:xsym2}
\end{equation}
The contraction of two CGTs $C \ast D$ necessarily yields
another CGT labeled $E$ in (\ref{fg:xsym2}b), with its own
orthonormal OM space. Assuming its OM space is complete, then
the contraction of $C\ast D$ can be projected by inserting the
identity $E^\dagger E = 1$ in (\ref{fg:xsym2}b). When fully
contracting the conjugate $E^\dagger$ onto $C\ast D$, this
results in what is referred to as an {\it X-symbol},
\begin{eqnarray}
   X ^{\mu\nu}_{\kappa} \equiv \mathrm{Tr} \Bigl[ \bigl(
      \mathrm{Tr}_{i_C^{(\ast)},i_D^{(\ast)}}[C^{\mu}\ast D^{\nu}] \bigr)
      \ast E^\dagger_{\kappa}
      \Bigr]
\text{ ,}\label{eq:xsym}
\end{eqnarray}
where $C$ is contracted on the subset of legs $i_C$ with $D$ on
legs $i_D$, the result of which is fully contracted with
$E^\dagger$, while keeping the OM indices $\mu$, $\nu$, and
$\kappa$ (green lines) open.

The X-symbol is derived from a contraction (hence `X') of a pair
of CGTs of arbitrary rank each and with fixed symmetry labels.
It has all TNS-related indices fully contracted [yellow lines in
(\ref{fg:xsym2}b)]. Consequently, X-symbols only operate in
between OM spaces with the three open indices $\mu$, $\nu$, and
$\kappa$. With $E$ also a CGT fixed by symmetry, it can be
referenced in the final object, again suggested by the shading
in (\ref{fg:xsym2}c). The overall result out of
(\ref{fg:xsym2}c) can now be inserted back into
(\ref{fg:xsym1}c).

The pictorial representation in (\ref{fg:xsym1}c) then
exemplifies the central result of this work: in the process of a
pairwise contraction of tensors, it suffices to contract the
centrally stored X-symbol onto the OM indices of the
corresponding pair of RMTs, thus merging $(\mu,\nu)$ and
decomposing ($\kappa$) the OM spaces. If the $w$ matrizes had
not been contracted onto the RMTs as in (\ref{fg:tensor4}c),
they can also be contracted here onto the X-symbol, instead.
Importantly, in the present context,
the summation over the OM indices $\mu$ and
$\nu$ in \Eq{eq:tensor:decomp} has turned into a 
regular contraction involving RMTs and X-symbols only. 
With the index $\kappa$ left open now, it is $E$ here that
can be simply referenced as indicated by the shading in
(\ref{fg:xsym1}c). Nothing else remains to be done on the level
of CGTs themselves. Therefore if all X-symbol are available and
up to date in the database, the CGTs themselves can be
completely side-stepped. The X-symbols fully take care of the
symmetry related multiplicity spaces in an efficient and general
manner.

As apparent from the definition in \Eq{eq:xsym}, each X-symbol
needs to remember where it came from via metadata. This includes
references to the three participating CGTs $C$, $D$, and $E$
(which also specifies all their symmetry labels, order and
direction of legs), and what indices have been contracted [$i_C$
and $i_D$ in \Eq{eq:xsym}]. Moreover, depending on the context,
one may have to contract the conjugate of the input tensors $C$
or $D$, instead [indicated then by $i_C^\ast$ or $i_D^\ast$ in
\Eq{eq:xsym}]. Hence the X-symbol also stores conjugation flags
for all three CGTs.
If the OM space is build successively via contractions as they
occur, the X-symbol also needs to remember identifiers as to the
state of CGTs such as a high-resolution time stamp of their last
modification time when the X-symbol was computed. Then if any
of the CGTs gets updated later along the course of a
calculation, these serve as flags as to whether or not also the
X-symbol needs to be updated when the same contraction is
encountered again at a later stage.

Note that if $C$ and $D$ already have complete OM individually,
this does {\it not at all} imply that also the $C\ast D$ will
exhaust the OM space of the resulting CGT $E$ [cf.
\Eq{eq:OM:dim}!]. When (re)computing an X-symbol, if $E$ is
already present e.g. from other earlier contractions, $C\ast D$
needs to projected onto it. If the OM space of the current $E$
was already complete, the projection can fully represent the
result. If the OM space of $E$ was not complete, then new OM
components out of $C\ast D$ may arise, which need to be
extracted and orthonormalized via Schmidt decomposition
(performed twice for numerical stability). As this extends the
OM space of the CGT $E$, finally, it needs to be updated
centrally.

\subsection{Relation to $6j$-symbols}

Consider the fully contracted tensor network of four CG3s 
with a total of six contracted lines in (\ref{fg:6j}a),
\begin{equation}
   \includegraphics[width=0.96\linewidth]{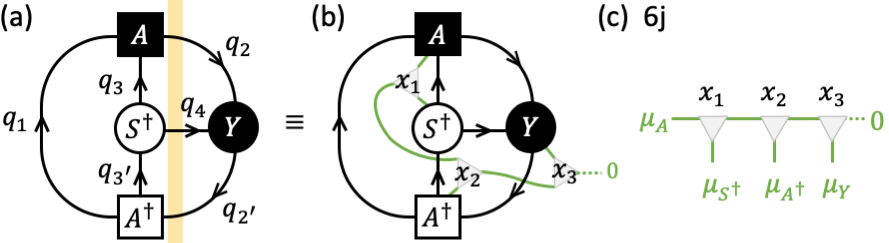}
\label{fg:6j}
\end{equation}
The part left of the yellow vertical marker, for example, may
represent the matrix elements of an operator $S^\dagger$ acting
on the local site incorporated by tensor $A$ in a matrix product
state \cite[see also \Eq{eq:Xconj}]{Schollwoeck11, Wb12_SUN,
Wb12_FDM}. With focus on CGTs here, the above tensor network is
viewed only at the level of CGTs in a particular configuration
with all $q$-labels fixed. Then the contraction of the TN left
of the yellow marker line results in a rank-3 CGT $(q_{2'} | q_2
q_4) \equiv (q_4 q_2 | q_{2'})^\dagger$. The result of the
contraction $A^\dagger \ast S^\dagger \ast A$ therefore
resembles the conjugate of a CG3.  In order to
determine the precise decomposition, it can be projected onto
the CGT $(q_4 q_2 | q_{2'})$, marked as $Y$ in (\ref{fg:6j}a,b).

The TN in (\ref{fg:6j}a) contains six contracted indices, and
therefore six symmetry sectors $q=\{ q_1, q_2, q_{2'}, q_3,
q_{3'}, q_4 \}$. They connect four CG3s, 
resulting in a fully contracted TN. Overall, therefore
(\ref{fg:6j}a) represents a `$6q$' symbol, or in the SU(2)
context, the well-known $6j$-symbol. Contrary to the case of
SU(2), however, which has no OM at the rank-3 level such that
$6j$ symbols are plain numbers, for general symmetries, each of
the four participating CGTs in (\ref{fg:6j}a) can carry outer
multiplicity. Therefore the resulting $6q$ symbol has four open
OM indices for a general non-abelian symmetry, i.e., represents
a rank-4 tensor purely in terms of OM indices.

Now the contraction in (\ref{fg:6j}a) can be performed pairwise.
The order of contractions is somewhat arbitrary, where the one
chosen in (\ref{fg:6j}b) is $(A^\dagger \ast (S^\dagger \ast A))
\ast Y$. Each of the pairwise contractions of CGTs can make use
of X-symbols. The first contraction $S^\dagger \ast A$, say,
makes use of the X-symbol $x_1$. The resulting CGT contracted
with $A^\dagger$ makes use of the X-symbol $x_2$, the result of
which when contracted with $Y$ makes use of $x_3$.
For the pairwise contraction of full tensors which include RMTs
and CGTs, the X-symbols actually need to be contracted onto the
RMTs as discussed with (\ref{fg:xsym1}c). In that sense, the TN
in (\ref{fg:6j}b) shows a TN in terms of the RMTs that come with
the CGTs in (\ref{fg:6j}a). Now the sequence of X-symbols
depicted in (\ref{fg:6j}b; green lines) can be isolated, thus
keeping OM indices open. With (\ref{fg:xsym2}c) then, the
resulting MPS-like sequence of X-symbols in (\ref{fg:6j}c)
exactly corresponds to the contraction of the CGTs in
(\ref{fg:6j}a).
Moreover, with the TN in (\ref{fg:6j}a) fully contracted, the
CGT describing the overall result is a rank-0 CGT, which itself
clearly has no outer multiplicity. Consequently, the sequence
in (\ref{fg:6j}c) stops with a singleton dimension indicated by
a dashed line with label `0'.

It follows therefore from the above constructive approach, that
any fully contracted TN built from CGTs can be decomposed into a
linear sequence of contractions based on X-symbols. In this
sense, X-symbols are equally general as $3n$-$j$ symbols, in
that any $3n$-$j$ symbol can be computed from them. Yet
X-symbols are much more naturally suited to tensor network
algorithms, in that they provide a general prescription for the
very elementary operation of a pairwise contraction of two
tensors of arbitrary rank each.

\subsection{Permutations and sorted CGTs\label{sec:perm}}

The indices of any tensor in a TNS need to be chosen in some
arbitrary but fixed order. The precise choice of order within
the TNS is typically a matter of convention, but of no further
concern, otherwise. Permutations, if performed correspond to
resorting of matrix elements according to the new index order.
The same also holds for CGTs. Therefore, by convention, it
suffices to only tabulate {\it sorted} CGTs which have their
symmetry labels ($q$-labels) sorted within the set of incoming
or outgoing indices, e.g., in a lexicographical style. Any
reference to a specific CGT then includes a reference to a
sorted CGT together with a permutation $p$ describing the actual
index order. The adopted sign convention is that the stored
sorted CGT starts with a positive coefficient. The permuted
references adhere to this original sorted tensor up to the
permutation only, i.e., there is no further sign adaptation
after permutation. This approach of making use of sorted CGTs,
i.e., with sorted $q$-labels, allows one to significantly reduce
redundancy of the CGTs that need to be stored, or subsequently,
also contracted.

The above is a well-defined prescription for CGTs that have
different $q$-labels on all of their legs. However, subtleties
arise if symmetry labels are degenerate, i.e., when precisely
the same representation occurs on more than one leg within the
group of either incoming or outgoing indices. Then the
prescription back and forth to sorted $q$-labels is not unique,
i.e., there can be a different permutation $p'$ that also leads
to the same sorted $q$-labels as the default permutation $p$
used in ones algorithm. In this case the permutation
$\mathtt{p}$ that transforms $p$ into $p'$ only operates within
degenerate $q$-label subspaces. It generates a non-trivial
orthogonal rotation $U_\mathtt{p}$ in OM space that needs to be
included in the permutation. This $U_\mathtt{p}$ can be
explicitly computed by fully contracting $C_q$ with the
conjugate of itself permuted by $\mathtt{p}$, the result of
which yields an X-symbol, say $X_\mathtt{p}$. As such, the
matrix $U_\mathtt{p}$ has all the properties of an X-symbol, and
hence can also be stored as such. Since $X_\mathtt{p}$
represents the full contraction of two CGTs, the resulting
rank-0 tensor cannot have OM. Therefore similar to the
discussion with (\ref{fg:6j}c), $X_\mathtt{p}$ has a singleton
trailing dimension, which can be skipped. This way, the X-symbol
$X_\mathtt{p}$ reduces to the matrix $U_\mathtt{p}$ above. It
can be absorbed into the matrix $w$ as in (\ref{fg:tensor4}b) if
$w$ is kept track of, or directly contracted onto the RMT for
the situation in (\ref{fg:tensor4}c).

\subsection{Generating standard rank-3 CGTs \label{Sec:triCGT}}

An elementary starting point for TN calculations are the
standard rank-3 CGTs (CG3s) that fuse the typically small
representations of a physical site. As a TN grows, however, the
multiplets on the auxiliary bond indices can quickly explore a
far larger set of representations. This implies for
contractions that often one does not require the full
tensor-product decomposition for every single rank-3 CGT
encountered, but only very specific combinations. Since the
full tensor-product decomposition of the fusion of two large
multiplets can become prohibitively expensive for large
symmetries [cf. \App{app:rsym}], the question arises, to what
extent specific (standard) rank-3 CGTs can be obtained by other
means.

Being interested in some specific CGT $(q_1 q_2 | q_3)$, in a
TNS setting, the typical situation is such that at least one of
the legs belongs to either the local state space of a physical
site or to the ireps according to which irreducible operators
transform. All of these are typically multiplets of small
dimension. Therefore in TN simulations, in practice, one of the
ireps $q_i$ ($i=1,2,3$) in a rank-3 CGTs can be considered
small. Frequently, it may even refer to the defining
representation or its dual, i.e., a primitive CGT
\cite{Butler76a,Searle88}. One
simple strategy to compute such rank-3 CGTs with their full OM,
is to exploit the freedom that arrows can be reverted at will
(while also switching to dual ireps). Hence, for example,
the multiplets $q_i$
can be sorted according to their dimension $|q_i|$. Then the
full tensor-product decomposition may be performed by taking the
tensor product of the {\it smallest} two ireps, and subsequently
reverting arrows as needed making use of \onej-symbols.

Rank-3 CGTs also appear routinely as the
result out of contractions. The remainder of this section
therefore is dedicated to the question
to what extent this can be used more systematically to
generate new (larger) CG3s with full OM.
As this only concerns symmetries, RMTs are fully ignored
in the following discussion.

\subsubsection{Triangular CGT networks}

A minimal CGT network required to compute a CG3 
from the contraction of other CG3s consists 
of the three CG3s in the triangular configuration
as shown in (\ref{fg:cgtri}a),
\begin{equation}
   \includegraphics[width=0.92\linewidth]{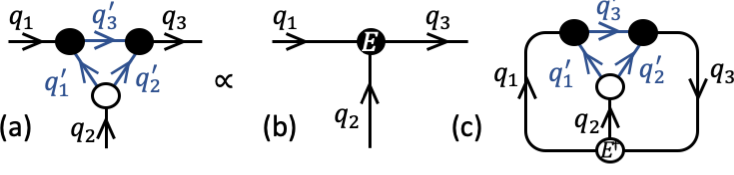}
\label{fg:cgtri}
\end{equation}
Once the closed lines along the triangle in (\ref{fg:cgtri}a)
are contracted, one obtains another CGT with three open legs.
In the present arrow configuration, the result is proportional
to a CG3 $(q_1 q_2|q_3)$ labeled $E$ in (\ref{fg:cgtri}b).
The proportionality factor is obtained by projecting $E$ onto
the result out of (\ref{fg:cgtri}a), as depicted in
(\ref{fg:cgtri}c). This situation then is completely analogous
to computing $6j$-symbol as discussed with (\ref{fg:6j}),
with the set of `$6j$'s $\{ q_1,q_2,q_3, q'_1, q'_2, q'_3\}$.
In general, $E$ itself can have outer multiplicity,
in which case the proportionality factor in (\ref{fg:cgtri}a)
becomes a matrix factor. Specifically, the
result out of contraction (\ref{fg:cgtri}a) needs to be
decomposed into the OM space of $E$ already present. If $E$ has
not yet been computed, the OM space arising out of
(\ref{fg:cgtri}a) needs to be orthonormalized (e.g., via QR
decomposition), thus defining the new $E$. If $E$ already
existed yet had not been obtained from a full tensor-product
decomposition, the contraction in (\ref{fg:cgtri}a) may yield
new OM components, and thus expands the OM space in $E$.

\subsubsection{Iterative schemes for CG3s with full OM}

New standard rank-3 CGTs, i.e., with two incoming and one
outgoing index, may be obtained systematically via a recursive
scheme based on contractions of smaller CG3s for the sake of
numerical efficiency
\cite{Schulten75,Butler76a,Searle88,Luscombe98}. To start with,
consider the CG3 $(q_1 q_2| q_3)$ in (\ref{fg:recurse}a),
\begin{equation}
   \includegraphics[width=0.96\linewidth]{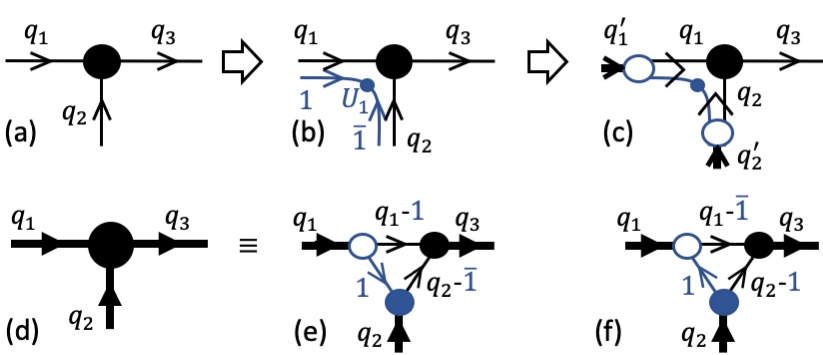}
\label{fg:recurse}
\end{equation}
Any non-abelian multiplet relevant in TN
simulations, by construction, is generated by building
a quantum-many body Hilbert space iteratively by adding
one particle after another
in the defining irep (if all possible multiplets can be
reached this way, then the representation is said to be
faithful \cite[Burnside theorem]{Butler76a}).
The number of particles relevant
in a given multiplet for SU(N) then relates to the number
of boxes. In this spirit, one can trivially
add a disconnected line in the defining irep
[labeled `1', e.g., one box in a SU(N) Young tableau]
along indices 1 and 2 as indicated by the
blue line in (\ref{fg:recurse}b). In order to have consistent
directions with $q_1$ and $q_2$, however, the arrow on the
blue line is reverted w.r.t. to say leg 2, which introduces the
\onej-symbol $U_1$ while also switching to the dual $\bar{1}$ on leg 2.
This way the particle entering on leg 1 is annihilated on leg 2,
and never affects leg 3.

The blue line can then be fused with both, $q_1$ and $q_2$
[(\ref{fg:recurse}c)], giving rise also to larger multiplets
$q'_{1}$ and $q'_{2}$ (indicated by thicker lines). The
contracted result in (\ref{fg:recurse}c) is again a CG3, 
but now in $(q'_{1},q'_2 | q_{3})$.
The \onej-symbol may be simply contracted onto any of the
connected CG3s (empty circles), thus resulting in a triangular
configuration similar to the one already 
encountered in (\ref{fg:cgtri}a).

The procedure above suggests that a CG3 
in some larger ireps $q'_1$ and $q'_2$ can be computed
via a contraction of three smaller CG3s (circles)
in a triangular TN that has smaller
ireps (thinner lines) on the contracted lines.
This also includes primitive CG3s,
i.e., CG3s that contain either the defining
irep $q_i=1$ or its dual $\bar{1}$ on one of their legs
\cite{Butler76a,Searle88}. Such CG3s [blue circles
in (\ref{fg:recurse})] are always OM free.
Starting in (\ref{fg:recurse}a) from a valid CG3, followed by
plain operations of adding and fusing a line in
(\ref{fg:recurse}b), the result in (\ref{fg:recurse}c) is
expected to be non-zero for any valid primitive CG3s
$(q'_1,1|q_1)$ and $(q'_2,\bar{1}|q_2)$, as long as 
they also result in a valid overall CG3 $(q'_1 q'_2|q_3)$.

The simple argument with (\ref{fg:recurse}a-c) can be
reformulated into a more general strategy on how to compute
larger CG3s recursively from smaller ones. Suppose one
is interested in computing the CG3 with larger ireps $(q_1
q_2|q_3)$, as depicted by thicker lines in (\ref{fg:recurse}d),
meaning that they are not the defining irep or its dual. Then one
may split off the defining irep from $q_1$, such that $(q_1| 1 ,
q'_{1})$ with $q'_{1} = q_1 -1 < q_{1}$. Here $q'_1$ is 
`smaller' as compared to $q_1$ in the sense that
$q'_{1}$ has one `fewer particle in the defining representation',
e.g., for SU(N) has one box fewer in the Young tableau
[note $q'_1$ does not necessarily have to be an irep
that is also smaller in dimension, since a
Young tableau with fewer boxes may represent a multiplet
of larger dimension].
Similarly, the dual of the defining representation, $\bar{1}$, is
split off from $q_2$, such that $(q_2| \bar{1} , q'_{2})$ with
$q'_{2} = q_2 -\bar{1} < q_{2}$ in the same sense as for $q'_1$
above, except that an `anti-particle' (hole) was removed from
$q_2$ where, e.g. for SU(N) the irep $\bar{1}$ consists of
$N-1$ boxes in a Young tableau, such that $N-1$ boxes are
split off from the Young tableau for $q_2$. Then merging the
\onej-symbol as in (\ref{fg:recurse}c) [blue dot] with the lower
CGT, this is equivalent to having a simple line in irep $1$
directed downward in (\ref{fg:recurse}e).
Depending on the external ireps,
one may also have to consider the reversed process
in (\ref{fg:recurse}f) where one particle moves upward.
For general non-abelian symmetry, typically also
multiple intermediate multiplets $q'_1$ and $q'_2$
can occur with the same number of `particles' or `boxes',
all of which need to be included, as this gives rise to
OM in the targeted $(q_1 q_2| q_3)$. Besides, the
CG3 $(q'_1 q'_2| q_3)$ itself may already have OM
which, however, is assumed to be complete from earlier
iterations, given that this CG3 is `smaller' w.r.t two
of its legs in the above sense.

Aiming at reducing the size of the constituting CG3s as in
(\ref{fg:recurse}e) fails for a maximum weight CG3,
having $q_3=q_1+q_2$. More generally, it fails for ireps
$q_3$ which have all particles or boxes present from $q_1+q_2$,
still. For these cases, $q_3$ can no longer be reached by
the smaller ireps $q'_1 = q_1-1$ and $q'_2=q_2-\bar{1}$. However, in the
present case this problem is simply taken care of by remembering
that directions on CGTs can be easily altered using
\onej-symbols. Therefore adopting a sorted order $q_1\ge q_2\ge
q_3$ (e.g. sorted lexicographically in terms of $q$-labels) this
excludes $(q_1,q_2|q_3)$ from ever becoming a maximum
weight CG3 for non-trivial $q_i>0$. As an aside,
note that for SU(N),
the multiplets $q_3$ out of $(q_1,q_2|q_3)$ all have the same
number of boxes in their Young tableau up to modulo $N$,
i.e., up to having full columns of $N=1+(N-1)$ boxes
removed, the latter corresponding to $1+\bar{1}$.
In (\ref{fg:recurse}e), by comparison, the maximum weight
state for $q'_1 = q_1-1$ and $q'_2=q_2-\bar{1}$ also has
$N$ fewer boxes, indeed, effectively
having removed one column in the Young tableau.

Consider still the specific example of SU(2) where,
for convenience and also consistency with general SU(N),
symmetry labels are taken as the integers $q_i\equiv 2S_i \geq 0$
for spin label $S_i$ \cite{Wb12_SUN}, as this also directly
specifies the number of boxes in the corresponding Yang tableau.
When computing the CG3 $(q_1 q_2|q_3)$ as in
(\ref{fg:recurse}e) then, with $1=\bar{1}$ for SU(2),
two fewer boxes in a Young-tableau can
reach $q_3$, i.e. $\delta q=2$, or
equivalently $\delta S=1$. Since the maximum weight CG3 was
excluded by reference to sorted CG3s above, this is no issue.
This shows that the decomposition in (\ref{fg:recurse}e),
indeed, works well for SU(2), in that two primitive CG3s (blue
circles) are contracted with a CG3 that strictly contains
smaller ireps on its legs. The present prescription is thus
analogous, e.g., to the 3-term recursive schemes for SU(2)
introduced in \cite{Schulten75,Luscombe98}
which also make use of primitive CG3s.

The strategy above to compute CG3s 
for specific $q$-labels via a recursive approach is general for
non-abelian symmetries from its outline. What is
explicitly demonstrated by construction in (\ref{fg:recurse}a-c),
suggests that (\ref{fg:recurse}e) permits to split off and
`route' a particle from leg 1 to 2 or vice versa (\ref{fg:recurse}f),
while it never reaches or participates in the multiplet
on leg 3 [cf. (\ref{fg:recurse}b)].
This approach above appears to work well empirically
to also generate the full OM space of the larger CG3 $(q_1
q_2|q_3)$ for arbitrary non-abelian symmetries. At the present
stage this remains a conjecture, though, and a rigorous
mathematical proof is left for the future.

In any case, explicit generation of all encountered CGTs becomes
prohibitive with increasing symmetry rank $\rsym{}$,
since typical multiplets grow exponentially in $\rsym{}$, in
practice like $\lesssim 10^{\rsym{}}$ [see \App{app:rsym}].
However these tensors are generated, even the explicit
evaluation of the primitive CG3s based on tensor product
decomposition fusing the defining irep only will quickly hit a
hard wall for $\rsym{}\gtrsim 5$ [e.g. with SU($N\gtrsim 6$)
already giving rise to a full tensor-product decomposition
of state spaces frequently exceeding several millions
in dimension, even if the problem is sparse at a density
of $\gtrsim 10^{-3}$].
On the other hand, given that the X-symbols introduced
in this paper represent fully contracted CGT networks
up to OM indices, this also makes
themselves susceptible to recursive build-up schemes.
This is an attractive route, since X-symbols 
are generally {\it much} smaller in dimension than the
contracted CGTs they represent. For the purposes of
this paper, however, this is left as an outlook.

\subsection{Summary}

This work introduces X-symbols for the efficient treatment of
pair-wise contractions in tensor networks in the presence of
general non-abelian symmetries. Once computed from CGTs and tabulated,
they permit to completely sidestep the explicit usage of CGTs at
a latter stage, as they are contracted onto the multiplicity
indices of the involved RMTs. X-symbols represent a general
framework that is also trivially applicable to abelian
symmetries. As such, they provide a coherent concept
for any type of symmetry setting.
Much of this paper is a summary of significant
extensions that have been implemented and already tested
thoroughly in the \QSpace v3 tensor library [see
\App{app:QSpace:v3}] with strong applications, e.g.,
in \cite{Liu15,Lee17dh,Chen18a,Wb18_SUN,Stadler19,
Walter19,Wang19}.
In this sense, the present paper provides
a concise, polished, and proven version of the underlying
concepts.


\begin{acknowledgments}

This work has greatly benefitted from a few intensive
discussions on non-abelian symmetries in general while visiting
Anthony Kennedy (University of Edinburgh, UK). I also
acknowledge general discussions and much appreciated support
from the group of Jan von Delft in Munich specifically also
including Seung-Sup Lee, Katharina Stadler, Benedikt Bruognolo,
Bin-Bin Chen, Jheng-Wei Li, and Arne Alex; furthermore Wei Li
(University of Beihang, China), Wang Yilin (Brookhaven National
Lab, US), Thomas Quella (University of Melbourne, Australia),
L\'aszl\'o Borda and Gergely Zar\'and (Budapest University,
Hungary). This work was support by the Heisenberg fellowship
(DFG WE4819/2-1), as well as by the U.S. Department of Energy,
Office of Basic Energy Sciences, under Contract No. DE-SC0012704
without temporal overlap.

\end{acknowledgments}

  \bibliography{../../Dropbox/source/info/mybib.bib}

\appendix

\section{Symmetry rank and typical multiplet dimension\label{app:rsym}}

The rank $\rsym{}$ of a symmetry is an intrinsic property of a
given simple non-abelian symmetry that needs to be
differentiated from the rank $r$ of tensors. The Lie algebra of
a simple non-abelian symmetry
possesses at most $\rsym{}$
simultaneously commuting generators which thus defines its rank.
This {\it Cartan subalgebra} $S_\alpha^{z}$ with
$\alpha=1,..,\rsym{}$ can be simultaneously diagonalized
\cite{Dynkin47,Cahn84,LiE92,Pope06,Gilmore06,Humphreys12},
and hence gives rise to an $\rsym{}$-dimensional label, or weight
space \cite{Wb12_SUN}. The Cartan subalgebra is complemented by
generalized raising and lowering operators
$S_\alpha^{(\dagger)}$ with $\alpha=1,..,\rsym{}$ which
correspond to the simple roots of the Lie algebra.

In practice, then the typical multiplet dimension encountered
grows exponentially with the rank $\rsym{}$ as in $|q| \sim
10^{\rsym{}}$ \cite{Wb12_SUN}.
The essential reason for this is
that multiplets explore an $\rsym{}$ dimensional volume of label
space derived from $S_\alpha^{z}$. This makes large symmetries,
such as SU(N) with $\rsym{SU(N)}=N-1$ for $N\gtrsim 4$
computationally challenging.
In contrast, the well-known SU(2) is a comparatively very simple
yet elementary non-abelian symmetry, in that it corresponds to a
symmetry of rank $\rsym{SU(2)}=1$. Therefore a single label
suffices (i.e., the spin $S$). For the same reason, there is no
inner multiplicity (IM) for SU(2), and also no outer
multiplicity (OM) for standard Clebsch Gordan coefficient
tensors (rank-3 CGTs).

\section{\QSpace v3 \label{app:QSpace:v3}}

The concepts for non-abelian symmetries presented in this paper
have been implemented over the past few years in the tensor
library \QSpace \cite{Wb12_SUN}. They were thoroughly tested in
a range of papers \cite{Wb07,Wb12_SUN,Wb12_FDM, Liu15, Lee17dh,
Chen18a,Wb18_SUN,Stadler19,Walter19,Wang19}
with many more applications via student and
other research projects. \QSpace was introduced in 2006 [in
hindsight as version 1 (v1)] on the level of arbitrary
combinations of abelian symmetries. This is effectively the
present state of the ITensor Library \cite{ITensor}. \QSpace v1
presented a convenient framework, with \citex{Wb07}
its first strong application.
\QSpace v2 \cite{Wb12_SUN}
was also able to handle non-abelian symmetries on a
generic level, by introducing an additional tensor layer for
generalized Clebsch Gordan coefficient tensors (CGTs). It built
a database for irreducible representations (R-store), as well as
CG3s \mbox{(C-store)}, i.e., with two arrows in
and one out. The C-store also stored a listing of the branching
rules out of each tensor product decomposition.

However, in \QSpace v2 the OM resolved CGTs were explicitly
attached to each tensor in full together with the RMTs.
Individual entries (fixed symmetry sectors) contained linear
superpositions in outer multiplicity, i.e., CGTs of type
$\sum_{\mu'} w_{\mu\mu'} C_{q\mu}$ [cf. discussion after
\Eq{eq:tensor:decomp}]. This turned out cumbersome since CGTs
were re-contracted in every \QSpace tensor contraction.
Specifically for larger symmetries, e.g., for SU($N\gtrsim4$),
this quickly shifted the dominant numerical cost from the actual
physical calculation w.r.t. the RMTs $\Vert A\Vert_q$ to the
treatment of the CGTs $C_q$.

Therefore for \QSpace v3 (developed and thoroughly tested since
2015), not just CG3s, 
but all CGTs of arbitrary rank are computed on demand once and
for all, properly orthonormalized [cf. \Sec{sec:CGT:norm}], and
stored in the C-store. The \QSpace tensor no longer carries
the full CGTs but only a reference. The X-symbols for pairwise
contractions are also computed on demand once and for all, and
stored in an additional database (X-store).

In \QSpace, databases are generally built on demand, except for
the very elementary initialization when a symmetry is used for
the very first time. Preemptive calculation of all possible
objects, e.g., up to some prespecified multiplet dimension
quickly proliferates to hundreds of thousand of entries,
otherwise, even though redundancy in storage has been minimized
to a large degree by using sorted CGTs, i.e., sorted w.r.t their
$q$-labels, etc. In the sense that it is impossible to build a
complete database for non-abelian symmetries that permit an
infinite number of ireps, it is mandatory eventually in any
case, to build entries on demand. While it would not really
matter for SU(2) or SU(3), since all objects are (re-)computed
quickly in these cases, for larger non-abelian symmetries, the concepts
of {\it on demand} and {\it once and for all} become crucially
important for numerical performance. For example, note that
starting with SU($N\geq 4$), the typical size of individual
multiplets in CGTs quickly surpasses the number of multiplets in
RMTs used in a calculation. For example, in DMRG simulations
one barely exceeds an effective dimension $D^\ast\lesssim 8,000$
within the space of RMTs. However, while typical multiplets for
SU(N) already reach dimensions up to $\lesssim 10^{N-1} = 1,000$
for SU(4), the largest generated multiplets there already reach
 $\lesssim 10^{N} = 10,000$ [cf. \App{app:rsym}].

Now building ones database on demand, quickly makes it dependent
on the history of a calculation. Even if CGTs are computed with
full OM, there can be an arbitrary orthogonal rotation in OM
space. In the absence of OM, this reduces to a simple sign
convention. However, for example, in SU(3), the CG3 
for $(qq|q)$ with $q=(nn)$ [using Dynkin labels;
\citenum{Dynkin47,Wb12_SUN}] with $n\geq 0$ has OM $m=n+1$, and hence can
already be made arbitrarily large. Using symmetries under
exchange of the 3-legs is of limited use here to fully
fix rotations in OM space. Given that larger-rank CGTs
are computed on demand via contractions, their specific rotation
in OM space necessarily will depend on the specific tensors
contracted first, and hence on the history of the calculation.
There exists no (at least to author) known convention that
naturally and fully fixes the basis of an OM decomposition in
all circumstances for CGTs of arbitrary rank.

Standard rank-3 CGTs (CG3s)
can be generated in two ways: (i) via
explicit tensor product decomposition (`standard Clebsch Gordan
coefficients'), or (ii) via contractions including
multiple rank-3 CGTs [cf. \Sec{Sec:triCGT}]
or a pair of tensors with rank $r>3$.
(i) is the clearly preferred option, since it
generates the full OM at once and in a deterministic and thus
well-defined manner. However, this can become prohibitive for
large symmetries, e.g., when considering the fusion of two large
multiplets on the bonds of a tensor network state,
while actually the full tensor-product decomposition
is not requires (e.g. since the large bond-multiplets
only fuse into the symmetry label of the global 
wave function which often is a singlet).
The full tensor-product of two large multiplets
may also be far too large to actually occur as a
multiplet at the level of a physical site in the lattice
Hamiltonian of interest. In this sense, there is (necessarily) a
cutoff in dimension: if a tensor product decomposition is
required, e.g., when building a product state space in TNS
setting, it needs to be performed in any case.
But if one needs to
compute a contraction that result in a rank-3 CGT, one may opt
to perform a full tensor-product decomposition first, and then
project the result of the contraction onto it. Or,
alternatively, one may be satisfied just with the result of the
contraction itself. If the resulting CGT already exists, the
result is projected onto it, thus possibly extending the
existing OM space. Either way may result in a different basis
in OM space. It is crucially important then, that one strictly
ensures consistency across ones calculations.

Now given a history dependent database that is accessed and
maintained centrally, this implies when running multiple jobs at
the same time or when parallelizing within a single job that
threads need to be coordinated. That is threads may have to
wait, if another thread is currently in the process of updating
the same object in the C-store (via contraction or
tensor-product decomposition) or the X-store (if a contraction
between a new pair of CGTs needs to be performed, or if a new OM
component was encountered such that the derived X-symbol needs
to be updated). This coordination can be enforced on the level
of the database (e.g. locks on affected objects) and also in
memory in between different threads (thread locks).

In summary, \QSpace v3 consists of three databases (referred
to as `RCX store' as a whole)
\begin{itemize}
\item R-store for irep representations generated in a calculation;
  these explicitly include a full basis decomposition in
  terms of weight labels, a sparse representation of the
  diagonal Cartan subalgebra (generalized $S_z$ matrizes)
  as well as of the simple roots of the Lie algebra
  (the generalized raising and lowering operators) under
  consideration [cf. \App{app:rsym}].
  
\item C-store for storage of all CGTs of arbitrary rank $r\geq 2$.
  The CGTs are stored in sparse format with typical average sparsity
  $\gtrsim 10^{-3}$. The sparse format necessarily requires a framework
  for sparse tensors of arbitrary rank
  which has been coded from scratch into \QSpace v2.
  The C-store also stores the fusion rules
  out of full tensor product decompositions, as well as all \onej-symbols,
  which simply represent a special case of CGTs, namely $(q\bar{q}|0)$.
\item X-store for the X-symbols that derive from any encountered
  pairwise CGT contraction that are {\it not} trivially zero due to
  non-permitted combinations of symmetry labels (e.g.
  when a CGT contraction were to result in a non-diagonal
  rank-2 CGTs, or a rank-3 CGTs that was not listed 
  in an earlier full tensor product decomposition).
\end{itemize}
The data in the R- and C-store is computed in better
than double precision (roughly quad), since the entire RCX store
is built iteratively along a TN calculation starting
from the very elementary defining representation
(and its dual, for convenience). This guards against
accumulated error and ensures that
all entries are numerically exact in double precision.
It is also important for sparse storage in order
to distinguish actual possibly small CGT coefficients
from numerical noise.
The X-symbols are computed from CGTs in the C-store,
but can eventually be cast into plain double precision
as they are contracted onto RMTs anyway.

The C-store for larger symmetries is extremely heterogeneous,
as it contains tensors that represent scalars, all the way
up to individual CGTs that [e.g. for rank-4 CGTs in SU(4) quickly]
require 1TB of space or larger. The X-store contains by far the
most of the entries. Many contractions are known to be trivially
zero since the symmetry labels of the resulting CGT are not
permitted from a symmetry point of view, and hence can
be excluded from the X-store. Still, e.g., by not insisting that OM
spaces are complete, the X-store also contains many
X-symbols that are actually zero, meaning that the pairwise
contractions of two CGTs results in a CGT of finite dimensions,
yet with (Frobenius) norm resembling numerical noise.

When running multiple jobs, it is convenient to maintain
a central global RCX-store that has strictly read-only access
(except for times when it is updated manually)
which contains the bulk of all symmetry related data.
In addition, a differential store that is local to each job,
allows each job to compute and store whatever is needed in addition.
Since the latter is decoupled from other running jobs, at least
at this level interference between different simulations
leading to possible inconsistencies
is avoided. Once an RCX-store is complete for a given
calculation (e.g., if the same calculation is run a second time),
only meta-data is read from the R- and C-store
(such as branching rules in tensor products or irep dimensions).
For contractions, only the required X-symbols in the X-store need to
be read once, with no need to explicitly load the full CGTs from
the C-store. The X-symbols are typically much smaller than
the involved CGTs, much like 6j symbols.

Since any tensor is stored as a \QSpace with the data
comprised as the tensor product in \Eq{eq:tensor:decomp:0},
in principle, it has access to {\it all}
matrix elements in the full state space. This makes \QSpace
tensors versatile \cite{Wb12_SUN}
in that all of the elementary tensor operations
are allowed that one is used to when performing calculations
{\it without} symmetries , as long as they do not explicitly break
a symmetry. For example, it is very difficult (since inconsistent)
to represent a finite magnetic field $B S_z$ if the calculation
was initialized with SU(2) spin symmetry. A representation of
$S_z$ would require to break up CGTs into specific components
which when preserving symmetries, however, are considered
inseparable units. In the presence of spontaneous symmetry 
breaking of an otherwise symmetric Hamiltonian, finally, symmetries
can be turned on and off at will. In particular, non-abelian
symmetries can also be reduced to their abelian core.
This is a valuable approach to shed light on physical scenarios
where spontaneous symmetry breaking is weak or debated.

\end{document}